\documentclass[ste,twoside]{stefano}

\usepackage{amsmath}
\usepackage{amssymb}
\usepackage{graphics}
\usepackage{rotating}
\usepackage{cite}
\usepackage{color}

\def\makeheadbox{{%
\hbox to0pt{\vbox{\baselineskip=10dd\hrule\hbox
to\hsize{\vrule\kern3pt\vbox{\kern3pt \hbox{  {\sc tunnelling
through two barriers} } \hbox{ {\sc
{\color{blue}{dma}}[{\color{black}{imecc}}]{\color{red}{UniCamp}}
} \hspace*{10.4cm} {\color{blue}{$\boldsymbol{\Sigma \delta
\Lambda}$}} }
\kern3pt}\hfil\kern3pt\vrule}\hrule}%
\hss}}}

\def\0{\mbox{\tiny $0$}}
\def\1{\mbox{\tiny $1$}}
\def\2{\mbox{\tiny $2$}}
\def\3{\mbox{\tiny $3$}}
\def\4{\mbox{\tiny $4$}}
\def\5{\mbox{\tiny $5$}}
\def\6{\mbox{\tiny $6$}}
\def\7{\mbox{\tiny $7$}}
\def\8{\mbox{\tiny $8$}}
\def\9{\mbox{\tiny $9$}}
\def\R{\mbox{\tiny $R$}}
\def\T{\mbox{\tiny $T$}}
\def\mi{\mbox{\tiny $-$}}
\def\pl{\mbox{\tiny $+$}}

\begin{document}
%

\title{\Large TUNNELLING THROUGH TWO BARRIERS}

\author{
Stefano De Leo\inst{1}
\and Pietro P. Rotelli\inst{2} }

\institute{
Department of Applied Mathematics, State University of Campinas\\
PO Box 6065, SP 13083-970, Campinas, Brazil\\
{\em deleo@ime.unicamp.br}
\and
Department of Physics, INFN, University of Lecce\\
PO Box 193, 73100, Lecce, Italy\\
{\em rotelli@le.infn.it} }


\date{{\em August, 2004}  }

\abstract{Recent studies of the tunnelling through two opaque
barriers claim that the transit time is independent of the barrier
widths and of the separation distance between the barriers. We
observe, in contrast, that if multiple reflections are allowed for
correctly (infinite peaks) the transit time between the barriers
appears exactly as expected.}










\titlerunning{\sc tunnelling through two barriers}

\maketitle


\noindent
 It is a well known result that the tunnelling time,
calculated by using the stationary phase method (SPM)
approximation in the limit of an opaque barrier, is independent of
the barrier width\cite{REP}. Such a phenomenon, called the {\em
Hartman effect}\cite{HE} implies arbitrarily large velocities
inside the barrier. When this is obtained with the use of the
non-relativistic Schr\"odinger equation, which can be considered
exact only when the velocity of light is infinite, there is no
paradox involved, only perhaps some doubt about the relevance. One
can in particular question, in this context, the use of the
terminology ``super-luminal velocities''. However, since similar
results have more recently been obtained with the Dirac
equation\cite{DE1,DE2}, there is good reason to be perplexed and
to invoke further analysis.\\
\indent We address in this paper a variant of this subject which
connects closely to a recent paper upon above-barrier diffusion
\cite{ABPD}. In recent years, the Hartman analysis has been
extended to a potential model with two successive barriers
separated by a free propagation region. Again in the opaque limit
for both barriers, it has been observed that, far from resonances,
the tunnelling phase time depends neither upon the barrier widths
nor upon the distance between the barriers \cite{TB1,TB2}. Thus,
this result predicts, contrary to common sense,
 unbounded group velocities even in the free region between
the two barriers. This phenomenon, also valid for an arbitrary
number of barriers \cite{NB1,NB2}, is known as the
\emph{generalized Hartman effect}. We shall demonstrate that a
different analysis, which allows for multiple scattering between
the barriers and, consequently, the existence of multiple peaks,
alters this result. Indeed the generalized Hartman effect
represents an example of an ambiguity in the
use of the stationary phase method. \\
\indent The starting point of the analysis is the one-dimensional
Schr\"odinger equation
\begin{equation}
\label{seq}  \Psi_{xx}(k,x) =  \mbox{$\frac{2\, m}{\hbar^{\2}}$}
\,[ \, V(x) - E \,] \, \Psi(k,x) \, \, ,
\end{equation}
for a particle of mass $m$ in a double barrier potential,

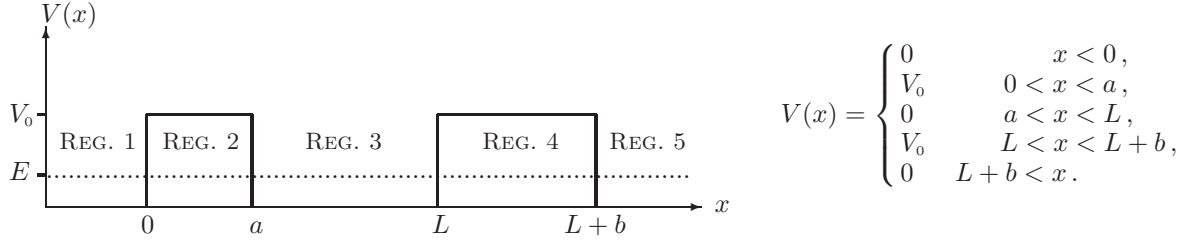
\begin{picture}(280,100) \thinlines
\put(2,10){\vector(0,1){68}} \put(0,80){$V(x)$}
\put(2,10){\vector(1,0){248}} \put(255,8){$x$}
\put(7,32){\mbox{\small \sc Reg. 1}} \put(46,32){\mbox{\small \sc
Reg. 2}} \put(100,32){\mbox{\small \sc Reg. 3}}
\put(167,32){\mbox{\small \sc Reg. 4}} \put(215,32){\mbox{\small
\sc Reg. 5}} \put(38,0){$0$} \put(79,0){$a$} \put(148,0){$L$}
\put(198,0){$L+b$} \thicklines \put(40,10){\line(0,1){35}}
\put(40,45){\line(1,0){40}} \put(80,45){\line(0,-1){35}}
\put(150,10){\line(0,1){35}} \put(150,45){\line(1,0){60}}
\put(210,45){\line(0,-1){35}}
\put(2,21){........................................................................................}
\put(-12,42){$V_{\0}$} \put(-1,45){\line(1,0){3}}
\put(-12,20){$E$} \put(-1,22){\line(1,0){3}} \put(280,42){$ V(x) =
\left\{
\begin{array}{lcl}
0 &  & \hspace*{.65cm} x < 0 \, ,\\
V_{\0} & ~~~~~ & 0 <  x <  a \, ,\\
0 &  & a < x < L\, ,\\
V_{\0} & ~~~~~~~  & \hspace*{-.05cm}L <  x <  L+ b \, ,\\
0 &  & \hspace*{-.65cm}L+ b < x \, .
\end{array}
\right.$}
\end{picture}

\noindent We have maintained different barrier widths because we
shall later discuss the case of an opaque limit only for the
second barrier but, for simplicity, we have given to the two
barriers the same height. It is easy to generalize our formulas to
the case of different barrier heights.

The standard procedure for finding the stationary solutions can
now be applied. The general solution for $\Psi(k,x)$ and any
$E<V_{\0}$ in the five regions are
\begin{equation}
 \Psi(k,x) \, = \,
\left\{
\begin{array}{lclcl}
 \mbox{\small \sc Region 1:} &~~~~ & e^{ikx} + A_{\1 \R} \,
 e^{\mi \, ikx} & &
 ~~~[\, k=\sqrt{2\, m \, E / \hbar^{\2}} \, \,]\, \, , \\
\mbox{\small \sc Region 2:} &  & \alpha_{\1} \, e^{\mi  \chi x} +
\beta_{\1} \,  e^{\,\chi x}  & &
~~~[\, \chi=\sqrt{2\,m\, (V_{\0} - E)/\hbar^{\2}} \, \,] \, \,,\\
\mbox{\small \sc Region 3:} & &A_{\1 \T} \, \left( e^{ikx} + A_{\2
\R} \,
 e^{\mi \, ikx} \right) \, \,, & &\\
 \mbox{\small \sc Region 4:} &  & A_{\1 \T} \, \left[ \alpha_{\2} \, e^{\mi \chi (x-L)} +
\beta_{\2} \,  e^{\,\chi (x-L)} \right]\, \,, & & \\
\mbox{\small \sc Region 5:} & & A_{\1 \T} \, \, A_{\2 \T} \,
 e^{ikx} \, \,. & &
\end{array}
\right.
\end{equation}
The requirement of continuity of $\Psi$ and $\Psi_{x}$ at
$r=0,\,a,\,L$ and $L+b$ give the matching conditions from which,
after some algebraic manipulations, we find
\begin{eqnarray}
\label{gco} A_{\1\R} &=&\mbox{$\frac{ \sinh{(\chi a)} }{
\sinh{(\,\chi a + 2 \,i \,\varphi\,)}}$}  \,\left[\,  1 -
\mbox{$\frac{ \sinh{(\,\chi b\,)} \, \sinh{(\,\chi a \,-\, 2\, i
\,\varphi\,)}}{\sinh{(\,\chi a \,)}\,\sinh{(\,\chi b \,+ \, 2 \,i
\,\varphi\,)}}$} \, \, e^{2 i
k ( L - a)} \, \right] \, / \, \mathcal{D} \nonumber \\
A_{\1\T} &=&\mbox{$\frac{2\, i\, \chi\, k}{w^{\2}}$} \,
\mbox{$\frac{e^{\mi i k a} }{\sinh{(\,\chi a + 2 \,i
\,\varphi\,)}}$}\, / \, \mathcal{D} \nonumber \\
A_{\1\T} A_{\2\R} & = & \mbox{$\frac{2\, i\, \chi\, k}{w^{\2}}$}
\, \mbox{$\frac{\sinh{(\,\chi b\,)}\, \,e^{ik( 2 L \mi a)}
}{\sinh{(\,\chi a + 2 \,i \,\varphi\,)} \,\sinh{(\,\chi b + 2 \,i
\,\varphi\,)}  }$}\, / \, \mathcal{D} \nonumber \\
 A_{\1\T} A_{\2\T} & = & \left(\mbox{$\frac{2\, i\,
\chi\, k}{w^{\2}}$}\right)^{\2} \, \mbox{$\frac{e^{\mi i k (a\pl
b)}}{\sinh{(\,\chi a + 2 \,i \,\varphi\,)}\,\sinh{(\,\chi b + 2
\,i \,\varphi\,)}}$}\,/ \, \mathcal{D}
\end{eqnarray}
where
\[
\mathcal{D} = \left[ \, 1 - \, \mbox{$\frac{\sinh{(\,\chi a)}\,
\sinh{(\,\chi b)}}{\sinh{(\,\chi a \,+ \, 2 \,i \,\varphi\,)}\,
\sinh{(\,\chi b \,+ \, 2 \,i \,\varphi\,)}}$} \, \, e^{ 2 i k ( L
\mi a)} \,\right]\, \, \,\,\,\mbox{and}\,\,\,\,\,
\varphi=\arctan[\, \chi/k\,]\,\,.
\]
If we consider the (double) opaque limit, $\chi a$ and   $\chi b
\gg 1 $, we reproduce (when $a=b$) the coefficients which have
been used to obtain the generalized Hartman
effect\cite{TB1,TB2,NB2},
\begin{eqnarray}
\label{oco}
 A_{\1\R}& \approx & \exp [\,-\, 2\, i \, \varphi \, ]
 \, \, ,\nonumber \\
 A_{\1\T}& \approx& \mbox{$\frac{2\, \chi\, k}{w^{\2}}$}\, \,
\exp [\,- \chi \,a - \, i \, k \,L \, ]\,\,/\,
\sin [\, 2\, \varphi - k\, (L-a)\, ]\, \, , \nonumber \\
 A_{\1\T}A_{\2\R}& \approx& \mbox{$\frac{2\, \chi\, k}{w^{\2}}$}\, \,
\exp [\,- \chi \,a + \, i \, k \,L  - 2 \,i \,\varphi \, ] \,\,/\,
\sin [\, 2\, \varphi - k\, (L-a)\, ]\, \, , \nonumber \\
 A_{\1\T}A_{\2\T}& \approx& \mbox{$\frac{8\, i\,
\chi^{\2}\, k^{\2}}{w^{\4}}$} \, \exp [\,- \chi \,(a+b) - \, i \,
k \,(L+b)  - 2\,i \,\varphi \, ] \,\,/\, \sin [\, 2\, \varphi -
k\, (L-a)\, ]\, \, .
\end{eqnarray}
We recall that all the above amplitudes are to be multiplied by
the plane-wave phase factor and the chosen modulation function
assumed peaked at $k = k_{\0}$. Without repeating all the
arguments leading to the generalized Hartman effect (including the
SPM itself), it suffices to note that, in the expression for
$A_{\1\T}A_{\2\T}$ of Eqs.(\ref{oco}), the $L+b$ dependence in the
phase will be cancelled by the plane wave phase calculated at
$x=L+b$. There remains an $L-a$ dependence in the modulus but that
does not affect the time for the appearance of the maximum in
region 5, under the hypothesis of a \emph{single peak }.

Previous analysis upon the generalized Hartman effect can be
criticized on several grounds. One objection is the standard use
in the through-going phase of the peak momentum $k_{\0}$ of the
incoming wave. The Hartman time for transit through a single
barrier then reads as $\tau_{\0}=2\,\varphi'(k_{\0})$, where the
prime stands for differentiation with respect to $k$. This may be
a good approximation for narrow barriers, but it is generally
wrong for opaque barriers because the peak or maximum value in the
outgoing momentum amplitude is not the same as that of the
modulation function (incoming amplitude). The only exception is if
the modulation function, which transforms the sum of plane waves
into wave packets, is so sharply {\em truncated} that only
momentum close to the incoming maximum are involved. What we like
to call the ``filter effect'' of a barrier preferentially allows
the higher momentum components to transit. So, in the opaque
limit, only the highest incoming momentum passes through. For
example, the group velocity in region 5, $\tilde{k}_{\0}/m$ , is
always higher than the incoming group velocity $k_{\0}/m$. A more
precise expression for the Hartman time is thus
$\tilde{\tau}_{\0}=2\,\varphi'(\tilde{k}_{\0})$.

Another principal criticism is the implicit assumption of a single
reflected and
transmitted wave packet. Let us list some objections:\\
 1) There is never any theoretical reason given for this
 assumption.\\
2) A natural alternative exists which still uses the SPM but
   involves multiple reflections with the ``expected'' transit
   times (see below).\\
3) The momentum distribution of the transmitted wave clearly
   displays multiple momentum peaks.\\
4) For technical reasons it is non trivial to  resolve numerically
   the twin barrier problems in the opaque limit in which formally
   the transmitted probability goes to zero. However,
   if, for example, only the second barrier is made opaque and
   the first is kept narrow, it is very easy to perform numerical
   calculations and see the multiple peaks in the reflected amplitude.
   Similarly for the foreward/backward flowing peaks in the
   intermediate region 3\cite{FN}.

Now for the alternative approach. Before going to the opaque
limit, the denominator factor $\mathcal{D}$ in Eqs.(\ref{gco}),
can be legitimately  expanded as a series in the numerator. In
particular, in region 5, we obtain the following transmitted
amplitude
\begin{eqnarray}
\label{ta}
 A_{\1\T} A_{\2\T} & =&  \left(\mbox{$\frac{2\, i\,
\chi\, k}{w^{\2}}$}\right)^{\2} \, \mbox{$\frac{e^{\mi i k (a\pl
b)}}{\sinh{(\,\chi a + 2 \,i \,\varphi\,)}\,\sinh{(\,\chi b + 2
\,i \,\varphi\,)}}$}\, \times \nonumber \\
 & & \sum_{\mbox{\tiny $n=1$}}^{\infty} \left( \, \mbox{$\frac{\sinh{(\,\chi a)}\,
\sinh{(\,\chi b)}}{\sinh{(\,\chi a \,+ \, 2 \,i \,\varphi\,)}\,
\sinh{(\,\chi b \,+ \, 2 \,i \,\varphi\,)}}$} \, \, e^{ 2 i k ( L
\mi a)}\, \right)^{\mbox{\tiny $n-1$}}\, \, .
\end{eqnarray}
An (infinite) sum of terms each of which can be analyzed with the
SPM to determine the position of their maxima. This is just the
procedure used in ref.\cite{ABPD}. The phase of the first
transmitted peak ($n=1$) is given by
\begin{equation}
\label{fm}
 \Phi_{\1}=k\,x-E\,t-k\,(a+b)-\arctan[\, \tan(2\,
\varphi) \coth(\chi\,a)\,] -\arctan[\, \tan(2\, \varphi)
\coth(\chi\,b)\,]\, \, .
\end{equation}
Consequently, the first transmitted peak appears at $x=L+b$ in
perfect accord with the expected ($L-a$ dependent) transit time.
The dependence of the subsequent phases ($n=2,3...$),
\begin{equation}
\Phi_n = \Phi_1 + (n-1) \left\{\, 2\,(L-a) -\arctan[\, \tan(2\,
\varphi) \coth(\chi\,a)\,] -\arctan[\, \tan(2\, \varphi)
\coth(\chi\,b)\,]\, \right\}\, \, ,
\end{equation}
contain extra  multiples of $\,2\,(L-a)$. In fact, in the opaque
limit
\begin{equation}
\Phi_n \approx k\,x-E\,t-k\,(a+b)-4\, \varphi + (n-1) \,[\,
2\,(L-a) -4\, \varphi\,]\, \, ,
\end{equation}
the filter effect of the first barrier guarantees that the group
velocity in region 3 is the same as that in region 5. Thus, the
successive exit times for the maxima predicted by the SPM are
proportional to multiples of the back/foreward travel times in the
inter-barrier region plus multiples of twice the Hartman time,
\begin{equation}
t_n= (m/\tilde{k}_{\0})\, [\, (2n-1)\,(L-a) + 2\,n\,
\tilde{\tau}_{\0})\, ]\,\,.
\end{equation}
The multiples of the Hartman time that appear above are a
consequence of  the multiple double-reflections.  The so-called
``delay-time'' in reflection in region 3 coincides with the
Hartman transit time in the opaque limit since the relevant
momentum used in the SPM for both is the upper limit
$\tilde{k}_{\0}$.

We have tested, but will not demonstrate here, that even though
the outgoing amplitudes are essentially independent of the shape
of the assumed modulation function (since they are dominated by
the exponential factors of the hyperbolic-sine functions) the SPM
still works surprisingly well. The only improvement we would
suggest, given the asymmetry in the momentum distribution, is the
use of the average momentum value in place of the peak momentum
value.

There is an objection that can be made to our analysis so far. We
have not demonstrated that the transmitted wave packets are not
overlapping and hence the treatment of each separately may  be
wrong. Indeed, we shall now argue that in the opaque limit the
outgoing wave packets {\em do} overlap substantially.
Nevertheless, we still claim that our treatment is more sensible
than that of a single-wave SPM analysis. To determine the size of
the outgoing wave packets, we employ the Heisenberg uncertainty
principle. In the opaque limit, the dominating part of the modulus
of each of the outgoing wave-packets is given by the filter term:
$\exp[-\chi\,(a+b)]$. This means that $\Delta \chi =1/(a+b)$.
Consequently, $\Delta k=\langle \chi / k \rangle \Delta \chi \ll
1/(a+b)$. It follows that $\Delta x \gg (a+b)$. Thus, in the
opaque limit, for any fixed value of the inter-barrier separation
$L-a$, the wave packet widths will eventually exceed the
separation between successive peaks. Now the successive wave peaks
of the outgoing amplitude are, in the opaque limit, almost equal
in magnitude. For example in the case when interference does not
shift these maxima (low interference) a resulting ``dragon-like''
structure emerges. The highest first maximum appears, as given in
Eq.(\ref{fm}), at the transit time in region 3 plus the {\em two}
Hartman times (because of two barriers). However, one must ask
what physical significance can be given to this maximum. The only
definition of position that is reasonable (for broad wave packets)
is the mean position of the wave packet and this appears at a very
different time from the maximum (first peak). It is the asymmetry
predicted by our interpretation in the transmitted wave train that
makes this question of mean vs. maximum relevant, and
distinguishes our procedure from a single wave analysis.

In conclusion, we have shown that the generalized Hartman effect
is the consequence of treating the outgoing wave as a single wave
packet. For us, this is not the natural choice as we have
explained above. Even when the multiple outgoing wave packets
overlap to the extent that they appear as a single wave packet,
the position of the maximum is far removed from the mean position
and is hence of little physical significance.


\begin{thebibliography}{99}
\bibitem{FN}
Examples and plots relative to these last two points for specific
parameters and modulation functions will be presented elsewhere.
\end{thebibliography}
\end{document}